\documentclass[aps,showpacs,showkeys,preprintnumbers,amsmath,amssymb,nofootinbib]{revtex4}

\usepackage{graphicx}
\usepackage{color}

\def \be  {\begin{equation}}
\def \ee  {\end{equation}}
\def \ee  {\end{equation}}
\def \bea {\begin{eqnarray}}
\def \eea {\end{eqnarray}}

\def \Tr  {\bf{Tr}}

\begin{document}

%\preprint{ECTP-2012-04}

\title{Hadronic Equation of State and Speed of Sound in Thermal and Dense Medium}

\author{Abdel Nasser Tawfik\footnote{http://atawfik.net/}}
%\email{a.tawfik@eng.mti.edu.eg}
%\email{atawfik@cern.ch}
\affiliation{Egyptian Center for Theoretical Physics (ECTP), Modern University for Technology and Information (MTI), 11571 Cairo, Egypt}
\affiliation{World Laboratory for Cosmology And Particle Physics (WLCAPP), Cairo, Egypt}

\author{Hend Magdy}
\affiliation{World Laboratory for Cosmology And Particle Physics (WLCAPP), Cairo, Egypt}

\date{\today}

%%%%%%%%%%%%%%%%%%%%%%%%%%%%%%%%%%%%%%%%%%%%%%%%%%%%%%%%%%%%%%%%%%%%%%
%%%   Abstract
%%%%%%%%%%%%%%%%%%%%%%%%%%%%%%%%%%%%%%%%%%%%%%%%%%%%%%%%%%%%%%%%%%%%%%

\begin{abstract}

The equation of state  $p(\epsilon)$ and speed of sound squared $c_s^2$ are studied in grand canonical ensemble of all  hadron resonances having masses $\leq 2\,$GeV. This large ensemble is divided into strange and non-strange hadron resonances and furthermore to pionic, bosonic and femionic sectors. It is found that the pions represent the main contributors to $c_s^2$ and other thermodynamic quantities including the equation of state $p(\epsilon)$ at low temperatures. At high temperatures, the main contributions are added in by the massive hadron resonances. The speed of sound squared can be calculated from the derivative of pressure with respect to the energy density, $\partial p/\partial \epsilon$, or from the entropy-specific heat ratio, $s/c_v$. It is concluded that the physics of these two expressions is not necessarily identical. They are distinguishable below and above the critical temperature $T_c$. This behavior is observed at vanishing and finite chemical potential. At high temperatures, both expressions get very close to each other and both of them approach the asymptotic value, $1/3$. In the HRG results, which are only valid below $T_c$, the difference decreases with increasing the temperature and almost vanishes near $T_c$. It is concluded that the HRG model can very well reproduce the results of the lattice quantum chromodynamics (QCD) of $\partial p/\partial \epsilon$ and $s/c_v$, especially at finite chemical potential. In light of this, energy fluctuations and other collective phenomena associated with the specific heat might be present in the HRG model. At fixed temperatures, it is found that $c_s^2$ is not sensitive to the chemical potential.  

\end{abstract}


\maketitle

%%%%%%%%%%%%%%%%%%%%%%%%%%%%%%%%%%%%%%%%%%%%%%%%%%%%%%%%%%%%%%%%%%%%%%
%%%   Section I
%%%%%%%%%%%%%%%%%%%%%%%%%%%%%%%%%%%%%%%%%%%%%%%%%%%%%%%%%%%%%%%%%%%%%%

\section{Introduction}
\label{sec:intr}

Despite the outstanding understand of the structure of matter at energy density much larger than the critical value which defines the hadron-quark deconfinement phase transition, one of the yet-unsettled problems of theoretical physics is the characterization of equation(s) of state (EoS) describing the behaviour of thermodynamic quantities at finite temperatures and densities. On one hand, the lattice quantum chrormodynamics (QCD) are reliable, especially at very high temperatures and densities. On  other hand, the exact equation of state of the hadronic matter is still rather complicated. In describing the ground state properties of nuclear matter having a large number of finite nuclei, the Hartree-Fock theories using Skyrme effective interactions are shown to be quite successful \cite{hf13a,hf13b,hf13c}. Nevertheless, as concluded in \cite{shortcoming}, it seems that serious concerns about basic physical symmetries arise when using EoS derived from Skyrme interactions in framework of Hartree-Fock theories, especially at finite temperatures. Furthermore, it is found that the speed of sound seems to violate the causality constrains leading to superluminal phenomena.  Early lattice QCD calculations \cite{earlylqcda,earlylqcdb} have shown that the speed of sound possesses a {\it dip} in the critical region, where the deconfinement phase transition is believed to take place through a slow crossover. Such a dip becomes weak with refining the certainty of the lattice QCD calculations \cite{lqcd1,lqcd2,bazazev,lqcd3,lqcd4}. The refining is based - aiming others - on extreme enriching the computing facilities and utilizing powerful algorithms \cite{lqcd4}.

Various reasons speak for utilizing the physical resonance gas model (HRG) in predicting the hadron abundances and their thermodynamics. This model seems to provide a good description for the thermal  evolution of the thermodynamic quantities in the hadronic matter~\cite{Tawfik:2014eba,Karsch:2003vd,Karsch:2003zq,Redlich:2004gp,Tawfik:2004sw,Tawfik:2004vv,Tawfik:2006yq,Tawfik:2010uh,Tawfik:2010pt,Tawfik:2012zz} and has been successfully utilized to characterize the conditions deriving the chemical freeze-out at finite densities~\cite{Taw3b,Taw3c,Tawfik:2012si}. In light of this, the HRG model can be used in calculating the speed of sound using a grand canonical partition function of an ideal gas with all experimentally observed states up to a certain large mass as constituents. The HRG grand canonical ensemble includes two important features \cite{Tawfik:2004sw}; the kinetic energies and the summation over all degrees of freedom and energies of resonances. On other hand, it is known that the formation of resonances can only be achieved through strong interactions~\cite{Hagedorn:1965st}; {\it resonances (fireballs) are composed of lighter ones and so on}. In other words, the contributions of the hadrons and their resonances to the partition function are the same as that of collisionfree particles with some effective mass. At temperatures comparable to the resonance half-width, the effective mass approaches the physical one \cite{Tawfik:2004sw}. Thus, at high temperatures, the strong interactions are conjectured to be taken into consideration through including heavy resonances. It is found that the hadron resonances with masses up to $2\;$GeV are representing suitable constituents for the partition function ~\cite{Tawfik:2014eba,Karsch:2003vd,Karsch:2003zq,Redlich:2004gp,Tawfik:2004sw,Tawfik:2004vv,Tawfik:2006yq,Tawfik:2010uh,Tawfik:2010pt,Tawfik:2012zz}.
The restriction on hadron masses is suggested in order to avoid the Hagedorn singularity \cite{Hagedorn:1965st,hgdrn1a,hgdrn1b,Karsch:2003zq,Karsch:2003vd}, the limit in which the Hagedorn partition function no longer exists. In light of this, the validity of the HRG is limited to temperatures below the critical one, $T_c$.

Recently, the problematic of characterizing a hadronic equation of state has been discussed in Ref. \cite{satz2009}. In an ideal gas consisting of hadron resonances having Hagedorn mass spectrum \cite{Hagedorn:1965st}, the speed of sound has been calculated at different upper cut-off masses in the resonance mass integration. It is found that the speed of sound initially increases similarly to that of an ideal pion gas, until near $T_c$. Then, the hadron resonance effects seem to become dominant. This causes a vanishing speed of sound at $(T_c-T)^{1/4}$.

In present work, we introduce a systematic study for the  speed of sound squared based on the HRG model  ~\cite{Tawfik:2014eba,Karsch:2003vd,Karsch:2003zq,Redlich:2004gp,Tawfik:2004sw,Tawfik:2004vv,Tawfik:2006yq,Tawfik:2010uh,Tawfik:2010pt,Tawfik:2012zz}. In a grand canonical ensemble, the thermal and dense evolution based on HRG calculations is derived in section \ref{sec:mdl}. We distinguish between the bosonic and fermionic contributions to the thermodynamic quantities. Also, we distinguish between the contributions stemming from the pion gas and that from a gas consisting of all resonances with and without pions. Finally, we compare the results with the recent lattice QCD calculations \cite{lqcd4}.
 The HRG results and their comparison with full lattice QCD are given in section \ref{sec:rsl}. Section \ref{sec:conl} is devoted to the conclusions and outlook.

%%%%%%%%%%%%%%%%%%%%%%%%%%%%%%%%%%%%%%%%%%%%%%%%%%%%%%%%%%%%%%%%%%%%%%
%%%   Section II
%%%%%%%%%%%%%%%%%%%%%%%%%%%%%%%%%%%%%%%%%%%%%%%%%%%%%%%%%%%%%%%%%%%%%%

\section{Speed of Sound Squared in Hadron Resonance Gas}
\label{sec:mdl}

In the context of Special or General Relativity, the {\it barotropic} equation of state of a perfect fluid reads \cite{cs2causality}
\bea\label{eq:eos1}
\frac{1}{c^2}\, \frac{\partial p}{\partial \epsilon} &=& \omega,
\eea
where the infinitesimal change in the pressure $p$ is proportional to that in the energy density $\varepsilon$ through the quantity $\omega\equiv c_s^2/c^2$. This proportionality can be very well determined in hadronic matter and quark-gluon plasma (QGP). Furthermore, it can be determined over the hadron-quark phase confinement-deconfinement transition. The barotropic dependence of pressure is to be deduced from the lattice QCD simulations \cite{lqcd4}, using tree-level improved Symanzik gauge action and a stout smeared staggered fermionic action and temporal lattice size approaching the continuum limit. Along the line of constant physics, the masses of the quarks are set to their physical values.

\begin{figure}[htb!]
\centering{
\includegraphics[width=8.cm,angle=-90]{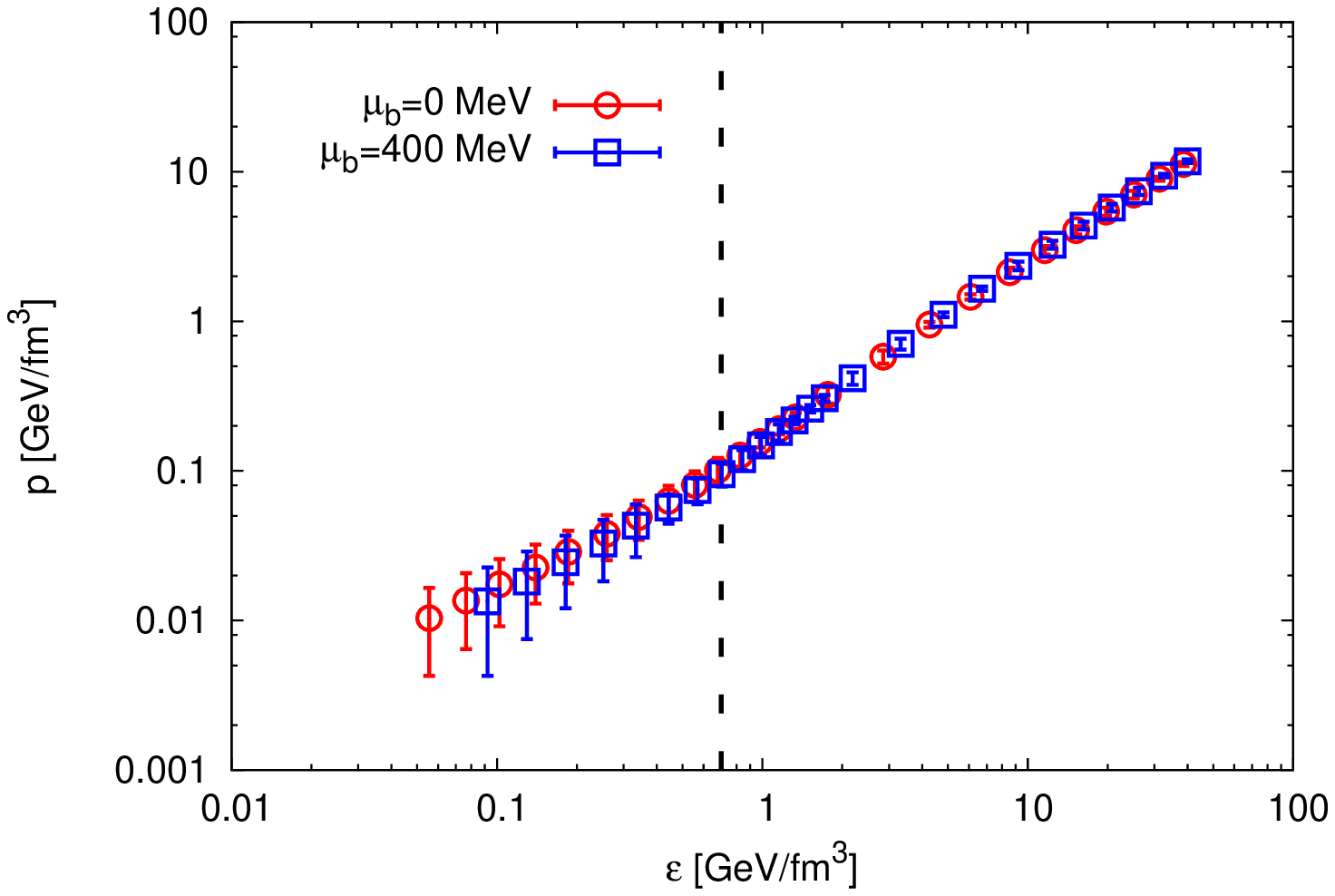} \\
\includegraphics[width=8.cm,angle=-90]{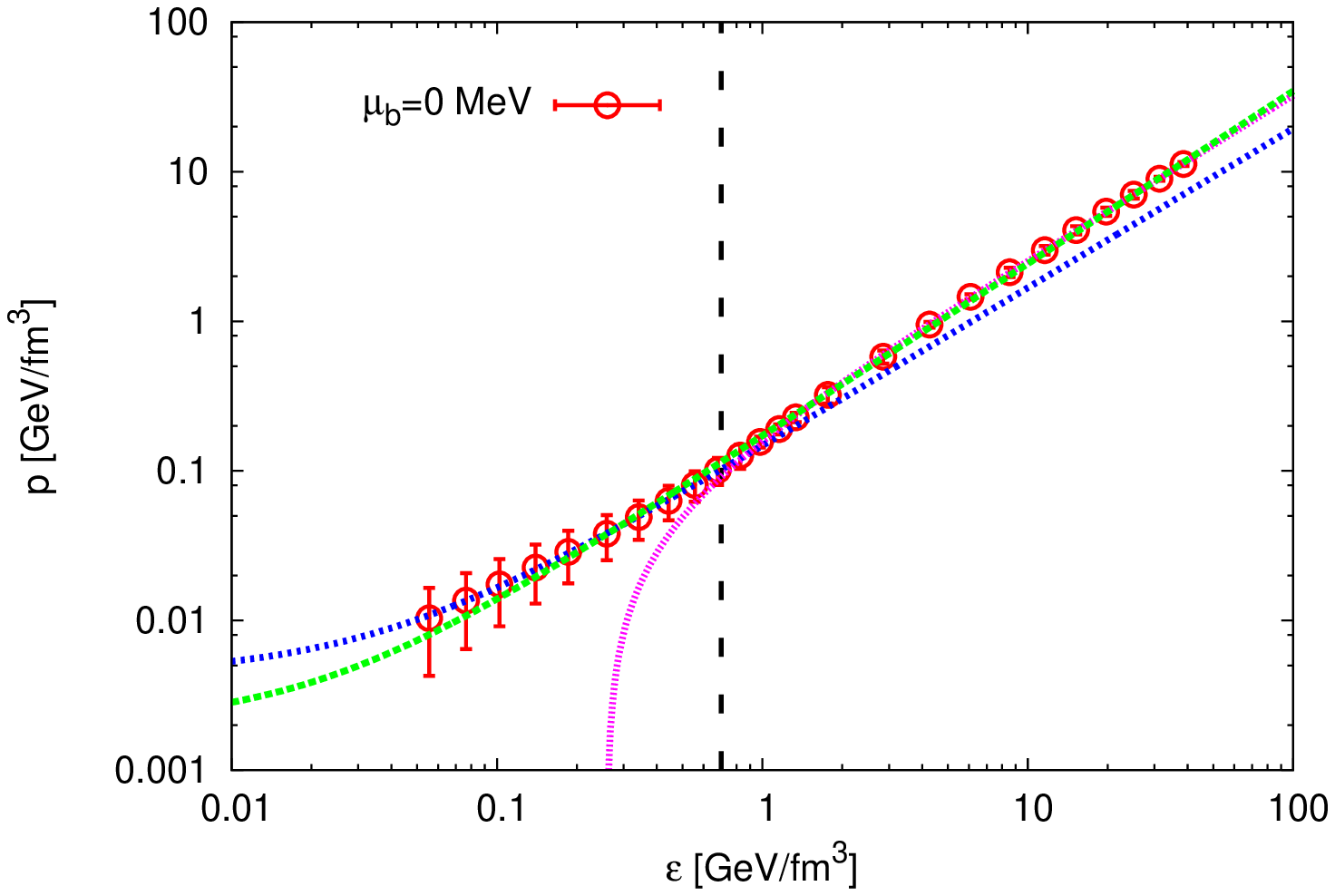}
\caption{The pressure density, $p$, is presented in dependence on the energy density, $\varepsilon$. Symbols are lattice QCD calculations \cite{lqcd4}. The top panel compares between the lattice QCD results at vanishing and finite baryon chemical potential, $\mu_b$. In bottom panel, the different phases  at $\mu_b=0~$MeV are fitted using Eq. (\ref{aeq:p1}).  The short-dashed and dotted curves stand for hadronic and partonic phase, respectively. The fitting of both phases is given by the dashed curve.  The small dip at $T_c$ seems to reflect the {\it slow} phase transition known as {\it crossover }.
\label{afig1a}
}}
\end{figure}

Top panel of Fig. \ref{afig1a} depicts EoS $p(\epsilon)$ in a wide range of temperatures, $T\leq 400~$MeV. It is obvious that the  dependence is almost linear referring to the nature of the phase transition from hadrons to quarks. This phase transition seems to be smooth, i.e. simply continuous and takes place over a considerable range of temperature. The nature of the phase diagram in lattice QCD has been discussed in~\cite{Tawfik:2004sw}. In the top panel of Fig. \ref{afig1a}, we present lattice QCD results at vanishing and finite baryon chemical potential ($\mu_b=400~$MeV). It is clear that the equation of state $p(\epsilon)$ is not sensitive to the change in the chemical potential. Therefore, we can analyse the results at a certain chemical potential, for instance $\mu_b=0~$MeV, and then generalize the results that the conclusions should be valid at arbitrary chemical potential. This is presented in the bottom panel. The dashed line represents the fitting in the entire $T$-region including hadronic and partonic phases. Ignoring the dip around $T_c$, the results can be fitted as a power law,
\begin{equation} \label{aeq:p1}
p(\varepsilon) = a_1 + a_2 \varepsilon^{a_3},
\end{equation}
where $a_1=0.002\pm 0.0023$, $a_2=0.17\pm 0.004$ and $a_3=1.158\pm 0.008$, respectively. As $a_1\rightarrow 0$ then $p(\varepsilon) \simeq a_2 \varepsilon^{a_3}$. In the hadronic phase, i.e. at temperatures $<T_c$, the previous power law seems to remain valid. Some changes appear in the parameters; $b_1=0.004\pm 0.001$, $b_2=0.173\pm 0.003$ and $b_3=1.066\pm 0.037$. Again, $p(\varepsilon) \simeq b_2 \varepsilon^{b_3}$. The results are given by short dashed line. In the quark phase, i.e. at temperatures $>T_c$, the lattice QCD results can be fitted at
$c_1=-0.045\pm 0.004$, $c_2=0.202\pm 0.003$ and $c_3=1.102\pm 0.005$. The results are given by dotted line.
Apparently, the expressions (\ref{aeq:p1})  imply that $c_s^2$ smoothly changes according to the changes in the phases.
\begin{itemize}
\item in the hadron-quark phase: $c_s^2=\partial p/\partial \varepsilon \simeq 0.17\; \varepsilon^{0.158}$, i.e. the speed of sound varies with the thermal evolution of the energy density. The latter has a non-monotonic behavior, when going from hadronic to partonic phases and vice versa. A recent review on lattice QCD calculations for the physical equation of state is given in Ref.   \cite{miller}. This can be seen when comparing
\item the hadronic phase, where $c_s^2\simeq 0.17\, \varepsilon^{1.07}$, with
\item the partonic phase, where $c_s^2\simeq -0.045 + 0.202\, \varepsilon^{1.1}$.
\end{itemize}
In light of this discussion, the effective EoS of the system of interest would be accessible through $c_s^2$.

The hadron resonances treated as a free gas~\cite{Tawfik:2014eba,Karsch:2003vd,Karsch:2003zq,Redlich:2004gp,Tawfik:2004sw,Taw3} are
conjectured to add to the thermodynamic pressure in the hadronic phase. This statement is valid for free as well as strong interactions between the resonances  themselves.
It has been shown that the thermodynamics of strongly interacting  system can also be approximated to an ideal gas composed of hadron resonances with masses $\le 2~$GeV ~\cite{Tawfik:2004sw,Vunog}. This cut is supposed to avoid the Hagedorn singularity \cite{hgdrn1a,hgdrn1b}. The grand canonical partition function reads
\bea
Z(T,V) &=&\Tr\left[ \exp^{-\mathbf{H}/T}\right],
\eea
where $\mathbf{H}$ is the Hamiltonian of such a system. The Hamiltonian is given by the sum of the kinetic energies of relativistic Fermi and Bose particles. The main motivation of using this Hamiltonian is that it contains all relevant degrees of freedom of confined and  strongly interacting matter. Implicitly it includes the interactions that result in resonance formation. In addition, it has been shown that this model can submit  a quite satisfactory description for the particle production in heavy-ion collisions. The comparison to the lattice QCD simulations has been introduced in Ref. \cite{Tawfik:2014eba,Karsch:2003vd,Karsch:2003zq,Redlich:2004gp,Tawfik:2004sw,Taw3}.

The dynamics of the partition function can be calculated exactly and be expressed as a sum over {\it single-particle partition} functions $Z_i^1$ of all hadrons and their resonances.
\bea\label{eq:lnz1}
\ln Z(T, \mu, V)&=&\sum_i^N \ln Z^1_i(T, \mu, V)=\sum_i^N\pm \frac{V g_i}{2\pi^2}\int_0^{\infty} k^2 dk \ln\left\{1\pm \exp\left[\frac{\mu_i -\varepsilon_i}{T}\right]\right\},
\eea
where $N$ is the total number of hadron resonances of interest, $\varepsilon_i(k)=(k^2+ m_i^2)^{1/2}$ is the $i-$th particle dispersion relation, $g_i$ is
spin-isospin degeneracy factor and $\pm$ stands for bosons and fermions, respectively.

Before the discovery of QCD, it was speculated about a possible phase transition of a massless pion gas to a new phase of matter. Based on statistical models like Hagedorn \cite{hgdrn1a,hgdrn1b} and Bootstrap \cite{boots1a,boots1b}, the thermodynamics of such an ideal pion gas is studied, extensively. After the QCD, the new phase of matter is known as QGP. The physical picture was that at $T_c$ the additional degrees of freedom carried by QGP are to be released resulting in an increase in the thermodynamic quantities. The success of HRG in reproducing lattice QCD results at various quark flavors and masses (below $T_c$) changed this physical picture, drastically. Instead of releasing additional degrees of freedom at $T>T_c$, it is found that the interacting system increases its effective degrees of freedom at $T<T_c$. In other words, the hadron gas has much more degrees of freedom than QGP \cite{Tawfik:2014eba,Karsch:2003vd}.

At finite temperature $T$ and chemical potential $\mu_i $, the pressure of $i$-th hadron or resonance reads
\begin{eqnarray}
\label{eq:lnz1}
p(T,\mu_i ) &=& \pm \sum_i^{N} \frac{g_i}{2\pi^2}T \int_{0}^{\infty}
           k^2 dk  \ln\left\{1 \pm \exp\left[\frac{\mu_i -\varepsilon_i}{T}\right]\right\}.
\end{eqnarray}
As no phase transition is conjectured in HRG, summing over all hadron resonances results in the final thermodynamic pressure in the hadronic phase. Switching between hadron and quark chemistry is given by the correspondence between  the {\it hadronic} chemical potentials and that of the quark constituents, for example,
$\mu_i =3\, n_b\, \mu_q + n_s\, \mu_S$, where $n_b$($n_s$) being baryon (strange) quantum number. The chemical potential assigned to the {\it degenerate} light quarks is $\mu_q=(\mu_u+\mu_d)/2$ and the one assigned to strange quark reads $\mu_S=\mu_q-\mu_s$. The strangeness chemical potential $\mu_S$ is calculated as a function of $T$ and $\mu_i $ under the assumption that the overall strange quantum number has to remain conserved in the heavy-ion collisions~\cite{Tawfik:2004sw}.  Based on this assumption, $\mu_S(\mu,T)$ is to be calculated at each value of the chemical potential $\mu$ and temperature $T$.

As given above, the speed of sound squared of single boson or fermion at finite  $T$  has to be calculated at fixed $s/n$, where $s$ and $n$ being entropy and number density.
\bea
c_s^2(T) &=& \left. \frac{\partial p(T)}{\partial \epsilon(T)}\right|_{s/n} = \left.\frac{d p(T)}{d T}\right|_{s/n} \; \left.\frac{d T}{d \epsilon(T)}\right|_{s/n}.
\eea
At finite baryon $\mu_b$ and strange chemical potential $\mu_S$. Assuming that $\mu=3\, n_b\, \mu_b + n_s\, \mu_s$, where $n_b (n_s)$ being baryon (strange) quantum number, then 
\bea \label{eq:cs2b}
c_s^2(T,\mu_i) &=& \frac{\frac{\partial p(T,\mu_i)}{\partial T} + \frac{\partial p(T,\mu_i)}{\partial \mu_i} \frac{d \mu_i}{d T}}{\frac{\partial \epsilon(T,\mu_i)}{\partial T} + \frac{\partial \epsilon(T,\mu_i)}{\partial \mu_i} \frac{d \mu_i}{d T}},
\eea
where the derivative $\partial p(T,\mu_i)/\partial \mu_i$ gives the number density, while 
\bea
\frac{\partial \epsilon(T,\mu_i)}{\partial T} &=& \pm \frac{g}{2 \pi^2} \frac{1}{T}\int_0^{\infty} k^2 dk\, \left[- \frac{\varepsilon\, e^{2\frac{\mu - \varepsilon}{T}}}{\left(1\pm e^{\frac{\mu - \varepsilon}{T}}\right)^2} \pm \frac{\varepsilon\, e^{\frac{\mu - \varepsilon}{T}}}{\left(1\pm e^{\frac{\mu - \varepsilon}{T}}\right)}\right]. \nonumber
\eea
The derivatives $d \mu/d T$ are to be evaluated under the two conditions that the ratio $s/n$ is kept fixed and the strangeness is conserved \cite{jeann}. In doing this, we distinguish between baryon and strange chemical potential and between baryonic and bosonic strange particles.
\bea
\frac{\partial p(T,\mu_i)}{\partial T} &=& \pm \frac{g_i}{2 \pi ^2}\frac{1}{T} \int _0^{\infty }k^2 dk\,\frac{(\varepsilon_i -\mu_i)}{1\pm e^{\frac{\varepsilon_i -\mu_i}{T}}} \pm \frac{g_i}{2
\pi ^2}\int _0^{\infty }k^2 dk\, \ln\left(1\pm e^{\frac{\mu_i-\varepsilon_i}{T}}\right) , \\
\frac{\partial \varepsilon(T,\mu_i)}{\partial T} &=& \pm \frac{g_i}{2 \pi ^2}\frac{1}{T}\int _0^{\infty }k^2 dk\,\frac{\varepsilon_i}{1\pm e^{\frac{\varepsilon_i -\mu_i}{T}}} - \frac{g_i}{2 \pi ^2}\frac{1}{T^2}\int
_0^{\infty }k^2 dk\, \frac{\varepsilon_i\,  e^{\frac{\mu_i}{T}}\, \left[(T-\varepsilon_i +\mu_i )e^{\frac{\varepsilon_i}{T}}\pm T e^{\frac{\mu_i}{T}}\right]}{\left(e^{\frac{\varepsilon_i
}{T}}\pm e^{\frac{\mu_i}{T}}\right)^2}, \label{eq:cv}
\eea
are entropy $s(T,\mu_i)$ and specific heat $c_v(T,\mu_i)$, respectively.
The second term of Eq. (\ref{eq:cv}) is complicated. It reflects different types of fluctuations related to the energy density and its product with the chemical potential
\bea
c_v(T,\mu_i) &=&
\pm \frac{g_i}{2 \pi ^2}\frac{1}{T}\int _0^{\infty }k^2 dk\,\frac{\varepsilon_i}{1\pm e^{\frac{\varepsilon_i -\mu_i}{T}}} \nonumber \\
&+& \frac{g_i}{2 \pi ^2}\frac{1}{T}\int _0^{\infty }k^2 dk\, \frac{\varepsilon_i\, e^{\frac{\varepsilon_i+\mu_i}{T}}}{\left(e^{\frac{\varepsilon_i}{T}}\pm e^{\frac{\mu_i}{T}}\right)^2} 
+\frac{g_i}{2 \pi ^2}\frac{1}{T^2}\int _0^{\infty }k^2 dk\, \frac{\varepsilon_i^2\, e^{\frac{\varepsilon_i+\mu_i}{T}}}{\left(e^{\frac{\varepsilon_i}{T}}\pm e^{\frac{\mu_i}{T}}\right)^2} 
-\frac{g_i}{2 \pi ^2}\frac{1}{T^2}\int _0^{\infty }k^2 dk\, \frac{\varepsilon_i\, \mu_i\, e^{\frac{\varepsilon_i+\mu_i}{T}}}{\left(e^{\frac{\varepsilon_i}{T}}\pm e^{\frac{\mu_i}{T}}\right)^2} \nonumber \\
&\mp & \frac{g_i}{2 \pi ^2}\frac{1}{T}\int _0^{\infty }k^2 dk\, \frac{\varepsilon_i\, e^{\frac{2\, \mu_i}{T}}}{\left(e^{\frac{\varepsilon_i}{T}} \pm e^{\frac{\mu_i}{T}}\right)^2}.  \label{eq:cvserieslong}
\eea
Such fluctuations are  related to the energy density and its product with the chemical potential.  
\begin{enumerate}
\item The last line of Eq. (\ref{eq:cvserieslong}) gives a direct relation to the {\it energy density} fluctuation, $\chi_{\varepsilon}$,
\bea
\mp \frac{g_i}{2 \pi ^2}\frac{1}{T}\int _0^{\infty }k^2 dk\, \frac{\varepsilon_i\, e^{\frac{2\, \mu_i}{T}}}{\left(e^{\frac{\varepsilon_i}{T}} \pm e^{\frac{\mu_i}{T}}\right)^2} 
&\equiv & \mp \frac{g_i}{2 \pi ^2}\frac{1}{T}\int _0^{\infty }k^2 dk\, \frac{\varepsilon_i\, e^{2\frac{\mu_i-\varepsilon_i}{T}}}{\left(1 \pm e^{\frac{\mu_i-\varepsilon_i}{T}}\right)^2}, \label{eq:cv4thterm}
\eea
which is nothing but $\langle \varepsilon\rangle \, T - \chi_{\varepsilon}$. 
\item The first line of Eq. (\ref{eq:cvserieslong}) gives the averaged energy density $\langle \varepsilon\rangle$. Appendix \ref{app:a} gives details on all these {\it energy density} dependent quantities. 
\end{enumerate}
Therefore, it becomes obvious that the second term in Eq. (\ref{eq:cv}) indeed reflects several types of energy fluctuations. Therefore, the speed of sound calculated through the ratio of entropy and specific heat, Eq. (\ref{eq:cs2a2}), seems to reflect essential dynamics and strong  correlations controlling the system under investigation. Furthermore, this might interpret the results given in section \ref{sec:rsl}, for instance , it would explain why the QCD barotropic pressure, the equation of state, is well reproduced by HRG, while $c_s^2$ is not.

Therefore, the speed of sound calculated from the ratio of entropy and specific heat, Eq. (\ref{eq:cs2a2}), seems to reflect essential dynamics and strong  correlations controlling the system under investigation. Furthermore, this might interpret the results given in section \ref{sec:rsl}, for instance , it would explain why $c_s^2=\left.\partial p/\partial \epsilon\right|_{s/n}$ and $c_s^2=s/c_v$ are not necessarily identical. In lattice QCD and HRG results, it is found that $c_s^2=\partial p/\partial \epsilon$ is larger than $c_s^2=s/c_v$, at $T<T_c$.  At $T>T_c$, $c_s^2=s/c_v$ gets larger than $c_s^2=\left.\partial p/\partial \epsilon\right|_{s/n}$.

%%%%%%%%%%%%%%%%%%%%%%%%%%%%%%%%%%%%%%%%%%%%%%%%%%%%%%%%%%%%%%%%%%%%%%
%%%   Section III
%%%%%%%%%%%%%%%%%%%%%%%%%%%%%%%%%%%%%%%%%%%%%%%%%%%%%%%%%%%%%%%%%%%%%%

\section{Results and Discussions}
\label{sec:rsl}

\begin{figure}[htb]

\includegraphics[angle=-90,width=8.cm]{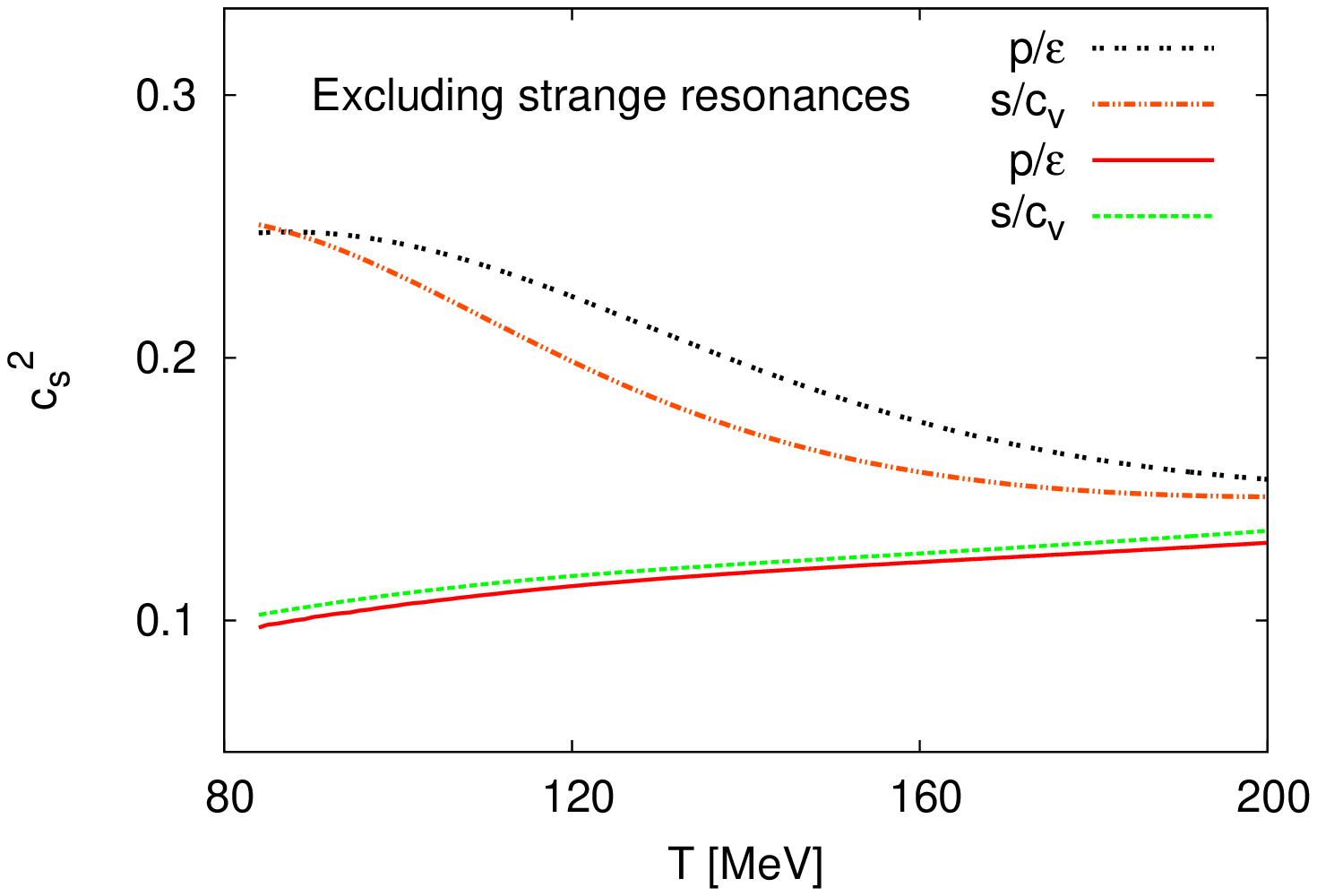}
\includegraphics[angle=-90,width=8.cm]{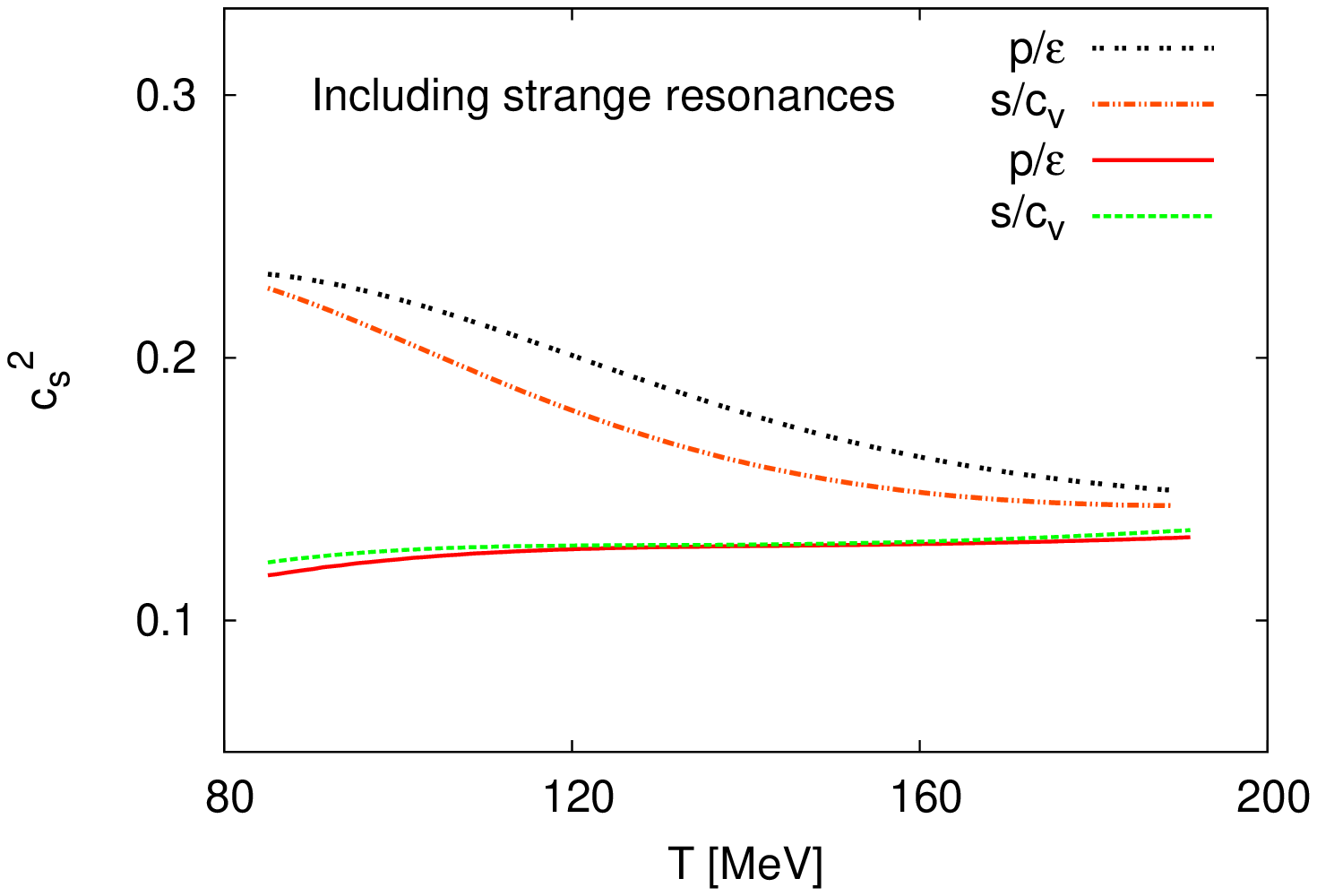}
\caption{Left panel: in HRG without strange hadron resonances, $c_s^2$ in presented in dependence on $T$ according to Eqs. (\ref{eq:cs2a1}) and (\ref{eq:cs2a2}).  The right panel draws the same dependence but with including the strange hadron resonances. In both graphs, bosonic (upper curves) and fermionic (bottom curves) sectors are utilized, separately.  \label{fig:woutqs1b}}
\end{figure}

In natural units, the thermal evolution of the speed of sound squared $c_s^2$  at finite chemical potential can approximately be given in two expressions, %Eq. (\ref{eq:cs24}) and (\ref{eq:cv}),
\bea
c_s^2(T,\mu) & = & \left.\frac{\partial\, p(T,\mu)}{\partial\, \varepsilon(T,\mu)}\right|_{s/n}, \label{eq:cs2a1}\\
c_s^2(T,\mu)  & = & \frac{s(T,\mu)}{c_v(T,\mu)}. \label{eq:cs2a2}
\eea
From mathematical point-of-view, these two expressions look identical. The physical nature should be deduced from the lattice QCD simulations \cite{lqcd3,lqcd4}. Based on this, the speed of sound squared seems to reflect not just a barotropic relation for the pressure, i.e. $p(\epsilon)$. The second expression relates $c_s^2$ to the entropy $s$ and the specific heat $c_v$. The latter encounters various fluctuations, especially in the energy density. This was given in the expression after Eq. (\ref{eq:cs2b}).

In Fig. \ref{fig:woutqs1b}, the thermal evolution of $c_s^2$ calculated in the HRG model using expressions (\ref{eq:cs2a1}) and (\ref{eq:cs2a2}) is depicted for boson (upper curves) and fermion (bottom curves) resonances, separately. In left panel, the strange resonances are excluded, while they are included in the right panel. We find considerable differences between boson and fermion hadron resonances. Also there is an obvious difference between the values of $c_s^2$ calculated with and without strange hadron resonances. Finally, the two expressions  (\ref{eq:cs2a1}) and (\ref{eq:cs2a2}) give distinguishable values of $c_s^2$, especially for bosons at the intermediate temperatures. It is apparent that the fermionic  $c_s^2$ is smaller than the bosonic one. The latter decreases faster than the increase taking place in the earlier.   Including strange hadron resonances increases the fermionic $c_s^2$, on one hand. On the other hand, it decreases the bosonic $c_s^2$.

In these calculations, the upper cut-off resonance mass is fixed at $\leq 2\,$GeV. Therefore, the  in/exclusion of strange resonances obviously changes the number of hadron resonances which are considered in the partition function, Eq. (\ref{eq:lnz1}).  It is expected that the values of thermodynamic quantities like pressure, energy density and entropy, are reduced when taking into account additional massive hadron resonances in the partition function. Apparently, this is not exactly the case with $c_s^2$, Fig. \ref{fig:woutqs1}. A reasonable explanation would be manifold. First, $c_s^2$ can be obtained from the ratio of two thermodynamic quantities, $p$ and $\varepsilon$ or $s$ and $c_v$. Second, an arbitrary change in $p$ and/or $s$ would not necessarily be the same as in $\varepsilon$ and/or $c_v$. In other words, any arbitrary change in denominator and numerator would not necessarily result in an change in the overall fraction. Finally, as introduced above $c_v$ likely associates with various fluctuations \cite{cvfluct}.

\subsection{The dominant role of pions at low temperatures}

\begin{figure}[htb]
\includegraphics[angle=-90,width=8.cm]{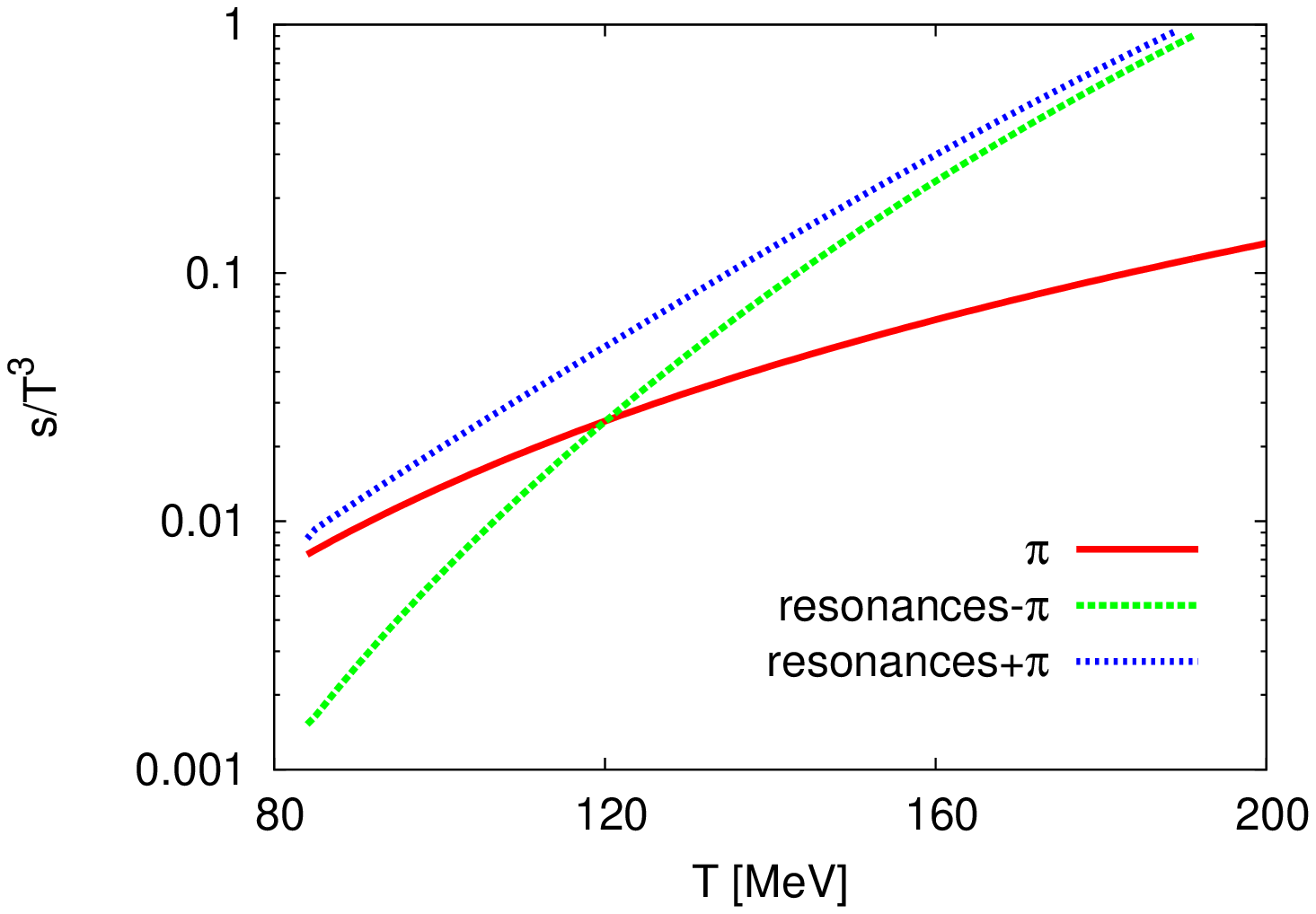}
\includegraphics[angle=-90,width=8.cm]{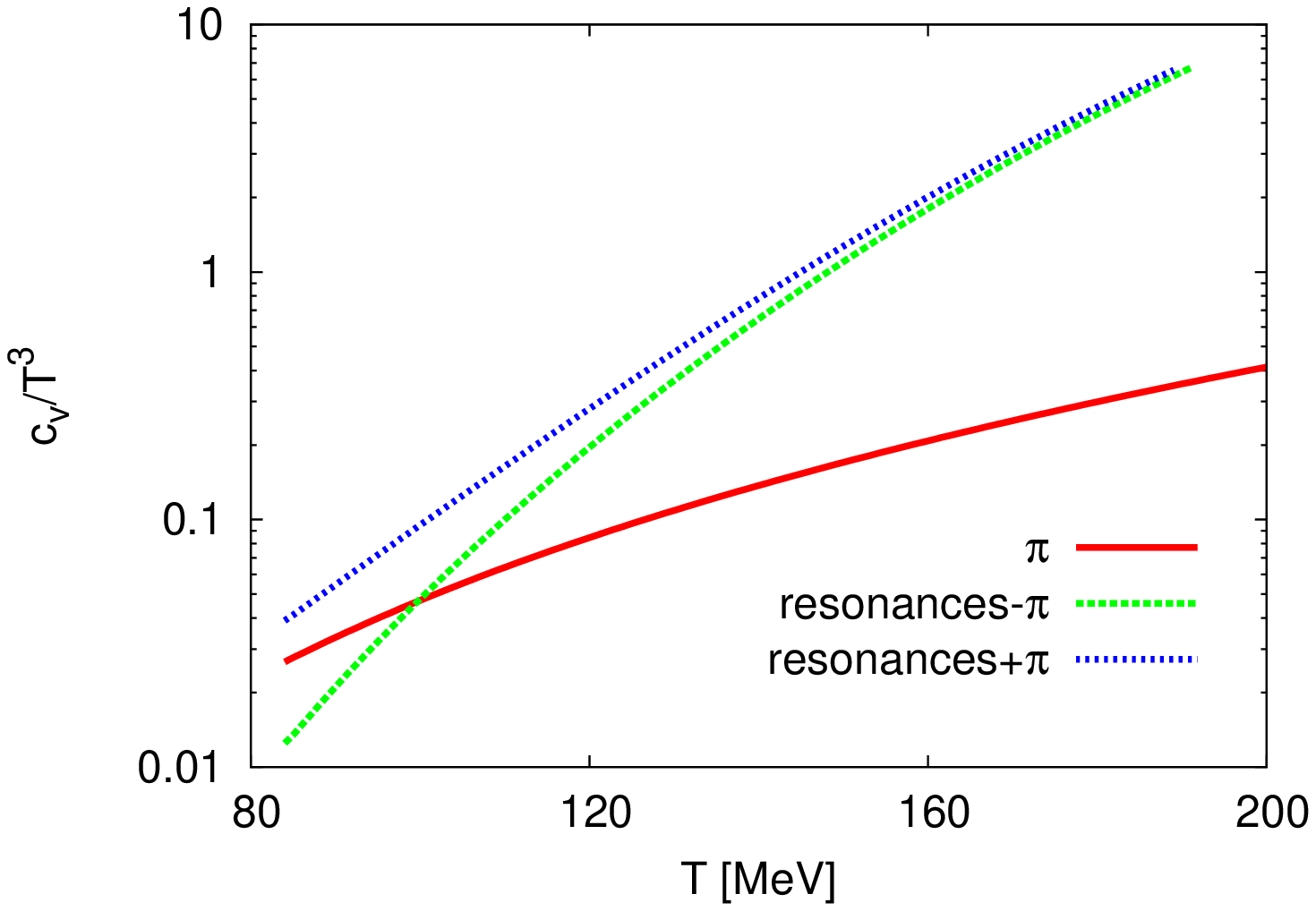}
\caption{The densities of dimensionless entropy $s/T^3$ (left panel) and $c_v/T^3$ (right panel) are given as functions of $T$ for pion gas (solid curve) and resonance gas (dashed curves), separately. Furthermore, the pions are included and excluded from the resonance gas.  \label{fig:woutqs1} }
\end{figure}

\begin{figure}[htb]
\includegraphics[angle=-90,width=8.cm]{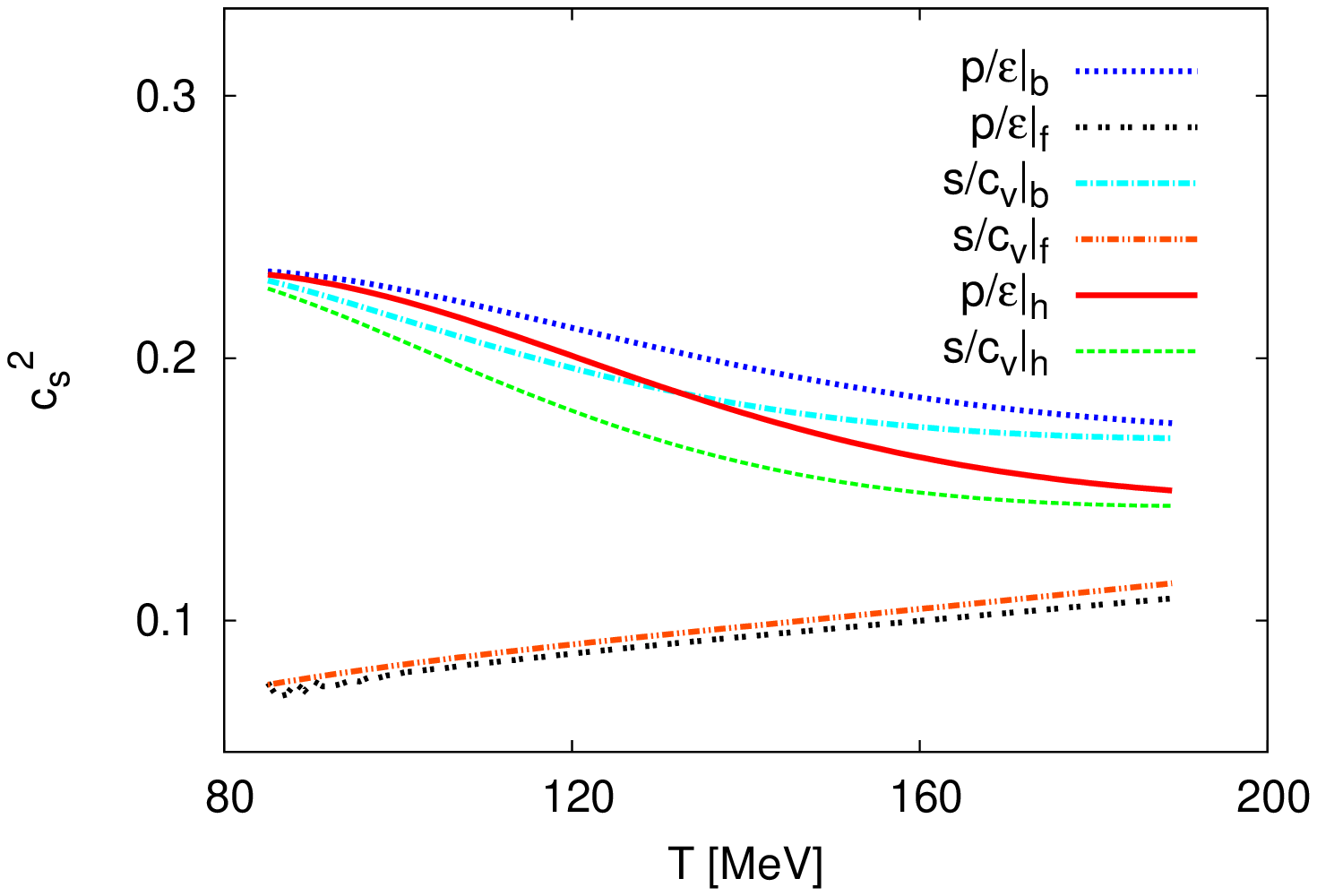}
\includegraphics[angle=-90,width=8.cm]{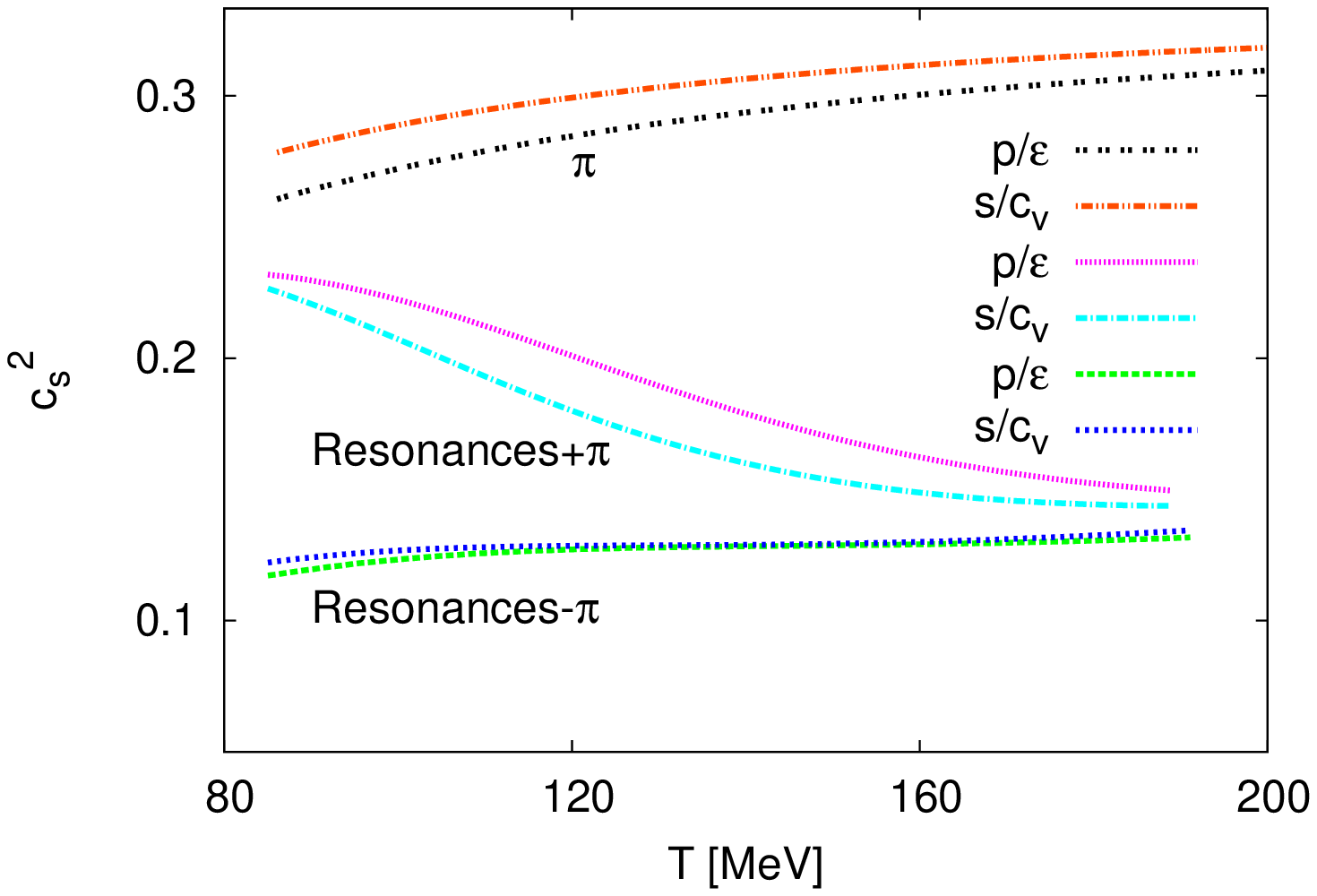}
\caption{Left panel: $c_s^2$ is calculated from the ratios of $p/\varepsilon$ and $s/c_v$ in the HRG model are given in dependence on $T$ for bosons (top lines), fermions (bottom lines) and hadrons (solid and dashed curves). A comparison between the values of $c_s^2$  calculated in the pion (top pair of curves) and the hadron resonance gas is drawn in the right panel. When pions are in/excluded from the hadron resonance gas, the resulting $c_s^2$  is illustrated as well (see text). 
\label{fig:pionshadrons1} }
\end{figure}

Bearing in mind the effect of adding massive resonance on the thermodynamics, we can expect that at low temperatures the pions likely plays the role of the effective mass. This effect has been illustrated in Ref. \cite{satz2009}, where $c_s^2$ calculated in the HRG model is found to increase similarly to that of an ideal pion gas. At low $T$, the heavy masses of the hadron resonances are minimally contributing  to the thermodynamics. At this $T$ scale, the lightest Goldstone bosons (pions) are dominant, as their masses are comparable to the $T$ scale, itself. This behavior is presented in Fig. \ref{fig:woutqs1}. In Fig. \ref{fig:pionshadrons1}, the thermodynamic quantities $s$ and $c_v$ are given in dependence on $T$ for different constituents, Eq. (\ref{eq:lnz1}). Here, we present another comparison. We distinguish between the contributions from pion and resonance gas. In other words, the pions replace all other resonances. They are excluded from the hadron resonance gas, on the other hand. It is apparent that  at low $T$ the pion gas gives much higher thermodynamic quantities (solid curves) than that of the resonance gas (dotted curves), so that when including pions in the resonance gas, the values remain almost unchanged. Up to $T\sim 120-140\,$MeV, the hadron resonances without pions gives thermodynamic quantities smaller than that from pions alone.  At higher $T$, the pions contributions are no longer dominant, so that the resonance gas with or without pions comes up with almost the same contribution.  The results of $c_v$ show that the dominance region of the pion gas is relatively short.

Absolving these preparations, successfully, then the thermal evolution of $c_s^2$ can be mapped out in the HRG model. In the left panel of Fig. \ref{fig:pionshadrons1}, we plot the same results given in the right panel of Fig. \ref{fig:woutqs1} aiming to differentiate between bosons and fermions. Both sectors are compared with the hadrons (solid and dashed curves). It is apparent that the bosons exclusively contribute to the peak at low $T$. Exacter said that the pions are the main contributors to the peak. At this temperature scale, the fermionic contributions are minimum. Increasing $T$ results in a decrease in the hadronic $c_s^2$ even below the one calculated in the bosonic gas, while the fermionic $c_s^2$ increases almost linearly and slowly.

Other features of this behavior can be revealed through the right panel of Fig. \ref{fig:pionshadrons1}. Here the comparison is made between hadronic and pionic $c_s^2$. The top pair of curves shows the pionic results. It is amazing that the values of $c_s^2$ are the largest. The bottom pair of curves represents the results from hadron resonances excluding the pions. Comparing the pions (the lightest Goldstone bosons) with the rest of hadron resonances (with masses up to $2\,$GeV) makes it clear that the values of $c_s^2$ are drastically reduced from $\sim 0.3$ for pions to $\sim 0.13$ for other resonances at $T>160\,$MeV. Adding the three pions to the hadron resonance results in the characteristic $c_s^2$ curves (middle curve). It starts with a peak, at low $T$. Increasing $T$ causes a slow decrease in $c_s^2$. At larger $T$, the decrease becomes fast, then becomes almost exponential. At these temperatures, the contributions of the three pions seem to be small, so that the results from the hadron resonances with and without pions become close to each other.

%%%%%%%%%%%%%%%%%%%%%%%%%%%%%%%%%%%%%%%%%%%%%%%%%%%%%%%%%%%%%%%%%%%%%%
%%%   Subsection IIIa
%%%%%%%%%%%%%%%%%%%%%%%%%%%%%%%%%%%%%%%%%%%%%%%%%%%%%%%%%%%%%%%%%%%%%%
\subsection{Confinement and Chiral Phase Transitions}

At vanishing light quark masses $m_q$, the broken chiral symmetry is assumed to be restored through  a  second order and in the $O(4)$ universality class \cite{refff7}. Defining the universal scaling allows to determine chiral pseudo-critical temperatures. This procedure is conjectured to remain applicable for $m_q \neq 0$, where chiral pseudo-critical temperatures are found to slightly deviate from the ones for $m_q=0$. For staggered fermions (partly preserve the chiral symmetry) at finite lattice spacing in the chiral limit, it is found that the relevant universality class is $O(2)$ rather than $O(4)$ \cite{refff8}. The numerical analysis shows that the differences between $O(2)$ and $O(4)$ universality classes are very small. 

The quark-antiquark chiral condensate can be used as an order parameter for the chiral transition. The derivative of pressure with respect to the constituent quark masses can be used to derive the light and strange chiral condensate \cite{Tawfik:2005qh} 
\begin{eqnarray}
\label{qqHRG}
<\bar{q}q>&=&<\bar{q}q>_0+ \sum_h \frac{\partial m_h}{\partial m_q}  \frac{\partial \Delta p}{\partial m_h},  \\ 
<\bar{s}s>&=&<\bar{s}s>_0+ \sum_h \, \frac{\partial m_h}{\partial m_s} \frac{\partial \Delta p}{\partial m_h}, 
\end{eqnarray}
where $m_h$ is the hadron mass and $<\bar{q}q>=<\bar{u}u>=<\bar{d}d>$ represents the light quark-antiquark condensate. $<\bar{q}q>_0$ and $<\bar{s}s>_0$ indicate the value of the light and strange quark-antiquark condensates in the vacuum, respectively. 
In free gas approximation, the contribution to pressure due to $m_h^{(i)}$, baryon charge $n_b^{(i)}$,  strangeness $n_s^{(i)}$, and degeneracy $g_i$ of $i$-th particle is given by
\begin{eqnarray}
  \label{p1}
  \Delta p_i=\frac{g_i\, (m_h^{(i)})^2\, T^2}{2\pi^2} \,
\sum_{n=1}^\infty \,\frac{(-\eta_i)^{n+1}}{n^2}
\,\exp\left(n\frac{n_b^{(i)}\, \mu_b  - n_s^{(i)}\, \mu_S}{T} \right)
\, K_2\left(n\frac{m_h^{(i)}}{T} \right),
\end{eqnarray}
where $\eta_i=\pm 1$ for fermions and bosons, respectively. $K_n(x)$ is the modified Bessel function.    

At finite temperature and chemical potential, the chiral  transition of two flavor QCD is analysed in the quark condensate, its dual and the dressed Polyakov loop with functional methods using a set of Dyson-Schwinger equations \cite{Fischer2011}. It is found that a  pseudo-critical temperature $T_{\chi}$ above the chiral transition in the crossover region but coinciding transition temperatures close to the critical endpoint. At vanishing chemical potential and for two quark flavors, $T_{\chi}=180\pm 5~$MeV \cite{Fischer2011}. 

The chiral transition is analysed in terms of universal $O(N)$ scaling functions with $2+1$ quark flavors. The extrapolations are based on simulations with different temporal extents with the HISQ/tree and asqtad action. The lattice QCD results on $T_{\chi}$ extrapolated to the continuum limit and the physical $m_q$ are presented in  \cite{Bazavov2012}.  The chiral transition temperature is $T_{\chi}=154\pm 9~$ MeV.

The $\mu_q$-dependence of the chiral phase transition temperature is parametrized as \cite{peter2012},
 \bea
 T_{\chi}(\mu_q) &=& T_{\chi}(\mu_q=0) - \kappa \left(\frac{\mu_q}{T_{\chi}(\mu_q=0)}\right)^2 + {\cal O}\left(\left(\frac{\mu_q}{T_{\chi}(\mu_q=0)}\right)^4\right),
 \eea
where $\kappa=0.059\pm 0.002$.

\subsection{Comparison with lattice QCD simulations at vanishing chemical potential}

\begin{figure}[htb]
\includegraphics[angle=-90,width=8.cm]{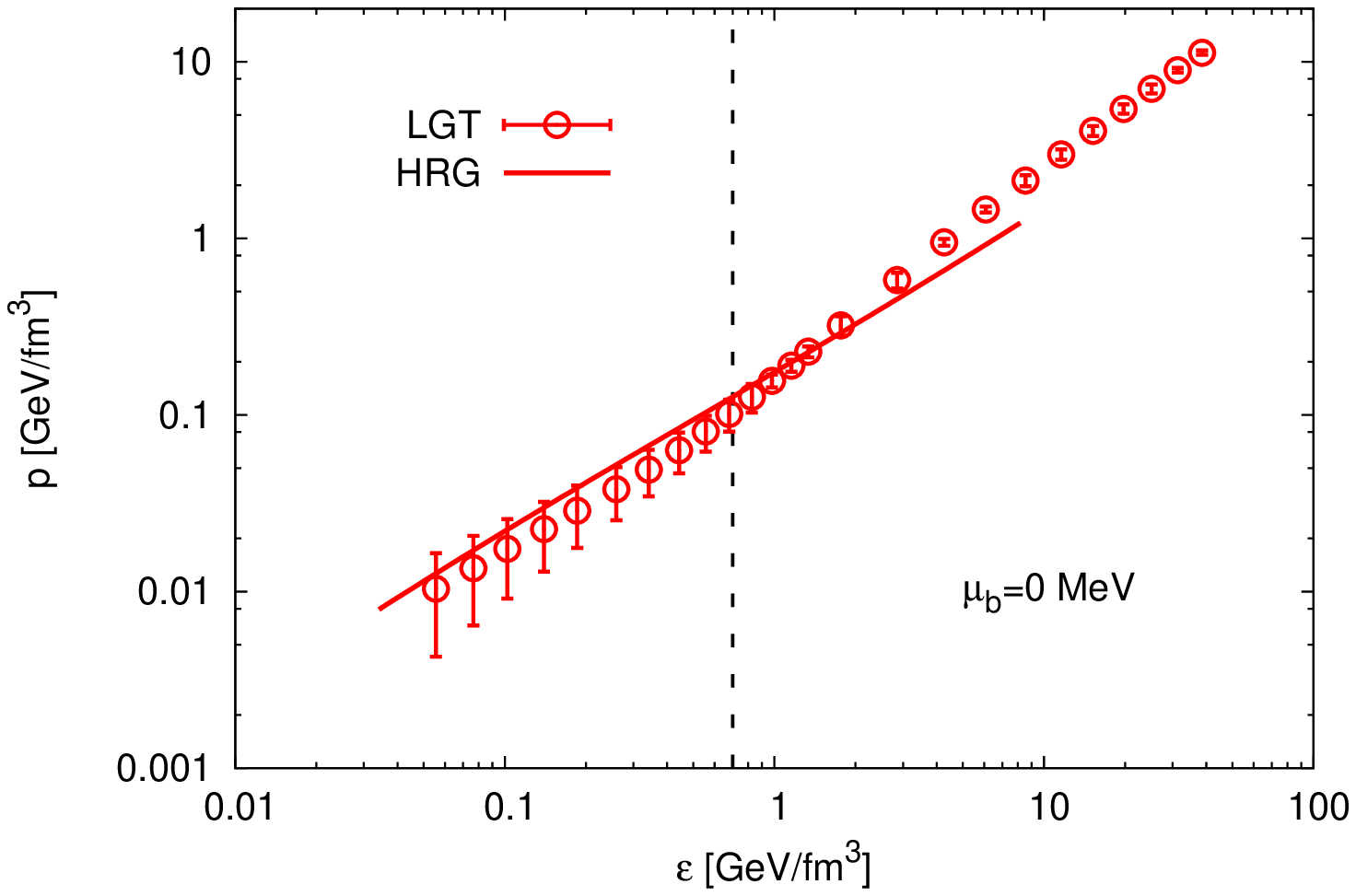}
\includegraphics[angle=-90,width=8.cm]{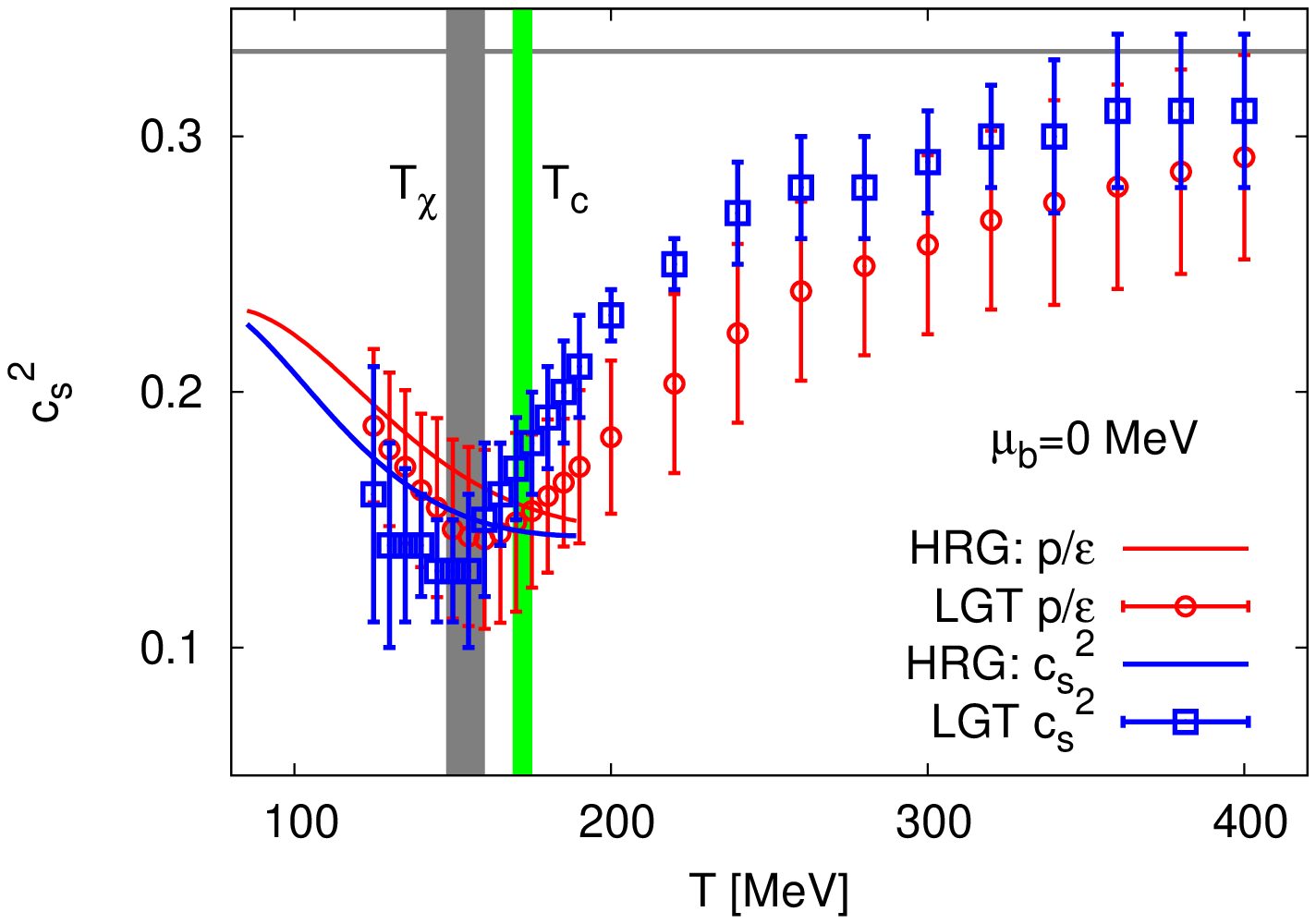}
\caption{Left panel: The barotropic dependence of the pressure $p$ calculated in lattice QCD \cite{lqcd4} (symbols) is compared with the HRG results (curves). Right panel: the thermal evolution of $c_s^2$  (symbols) calculated in lattice QCD is compared with the HRG results (curves). The critical line marks the critical transition. Beyond these marks, HRG is not longer valid.
\label{fig:cs2lqcd1} }
\end{figure}

The lattice QCD simulations \cite{lqcd4} introduce a systematic estimation for $c_s^2$ in thermal and dense medium. The equation of state of QCD for nonzero chemical potentials is determined via Taylor expansion of the pressure. Also here, $2+1$ quark flavors with physical masses on various lattice spacings are implemented.  In Fig. \ref{fig:cs2lqcd1}, the HRG results are confronted with these lattice QCD simulations \cite{lqcd4}.
In left panel in Fig. \ref{fig:cs2lqcd1}, the HRG barotropic dependence of the pressure is compared with the lattice QCD calculations at a vanishing baryon chemical potential. As given in Fig. \ref{afig1a} (top panel), the pressure has an almost linear dependence on the energy density, especially in hadronic and partonic phases, where the slopes would be used to determine corresponding $c_s^2$, directly.  The excellent agreement between HRG and lattice QCD in reproducing the thermodynamic quantities does not make it straightforward to find a {\it proper} interpretation for the small overestimation of $c_s^2$, as given in right panel of Fig. \ref{fig:cs2lqcd1}. On other hand, the lattice data have a minimum (dip). This is not reflected in the HRG results. A precise judgement about the dataset and its structure would be achievable through the normalized higher moments \cite{Tawfik:2012si}. In lattice QCD as well as in HRG, there is a qualitative tendency that $c_s^2$ slightly decrease with increasing $T$ up to $\sim 150\,$MeV. As noticed, the HRG model seems to overestimate $c_s^2$ calculated in lattice QCD. 
The deviations could be due to the pion ''doublers'' in the lattice QCD calculations, especially at low temperatures. The doublers effectively generate a larger ''mean pion mass'' and in a massive gas of the speed of sound squared.
On the other hand, there is another quantitative difference, this time between the HRG results themselves that are calculated by Eq. (\ref{eq:cs2a1}) and Eq. (\ref{eq:cs2a2}).

Regardless the quality of the comparison, it is clear that the difference between the thermal evolutions of $c_s^2$ calculated in HRG model using Eq. (\ref{eq:cs2a1}) and Eq. (\ref{eq:cs2a2}) seems to be approved by the lattice QCD simulations. In other words, $c_s^2$ calculated by Eq. (\ref{eq:cs2a1}) and Eq. (\ref{eq:cs2a2})  is apparently not identical. Both datasets show a remarkable difference. This would support the conclusion that the physics of calculating EoS via Eq. (\ref{eq:cs2a1}) and the physics of the fluctuations associated with the specific heat, Eq. (\ref{eq:cs2a2}).

At higher temperatures, $c_s^2$ calculated in lattice QCD raises, while in the HRG model it resumes its decreasing tendency. In this region of temperature, where the deconfinement phase transition is assumed to set on, there is a little quantitative and qualitative discrepancy. The lattice results increase, so that at $T\simeq 185\,$MeV, the HRG results are higher than that of the lattice QCD. The dip appearing in the lattice data is not present in HRG, at least in the hadronic phase, where HRG is exclusively applicable.

%%%%%%%%%%%%%%%%%%%%%%%%%%%%%%%%%%%%%%%%%%%%%%%%%%%%%%%%%%%%%%%%%%%%%%
%%%   Section IIIb
%%%%%%%%%%%%%%%%%%%%%%%%%%%%%%%%%%%%%%%%%%%%%%%%%%%%%%%%%%%%%%%%%%%%%%

\subsection{Comparison with lattice QCD simulations at finite chemical potential}

\begin{figure}[htb]
\includegraphics[angle=-90,width=8.cm]{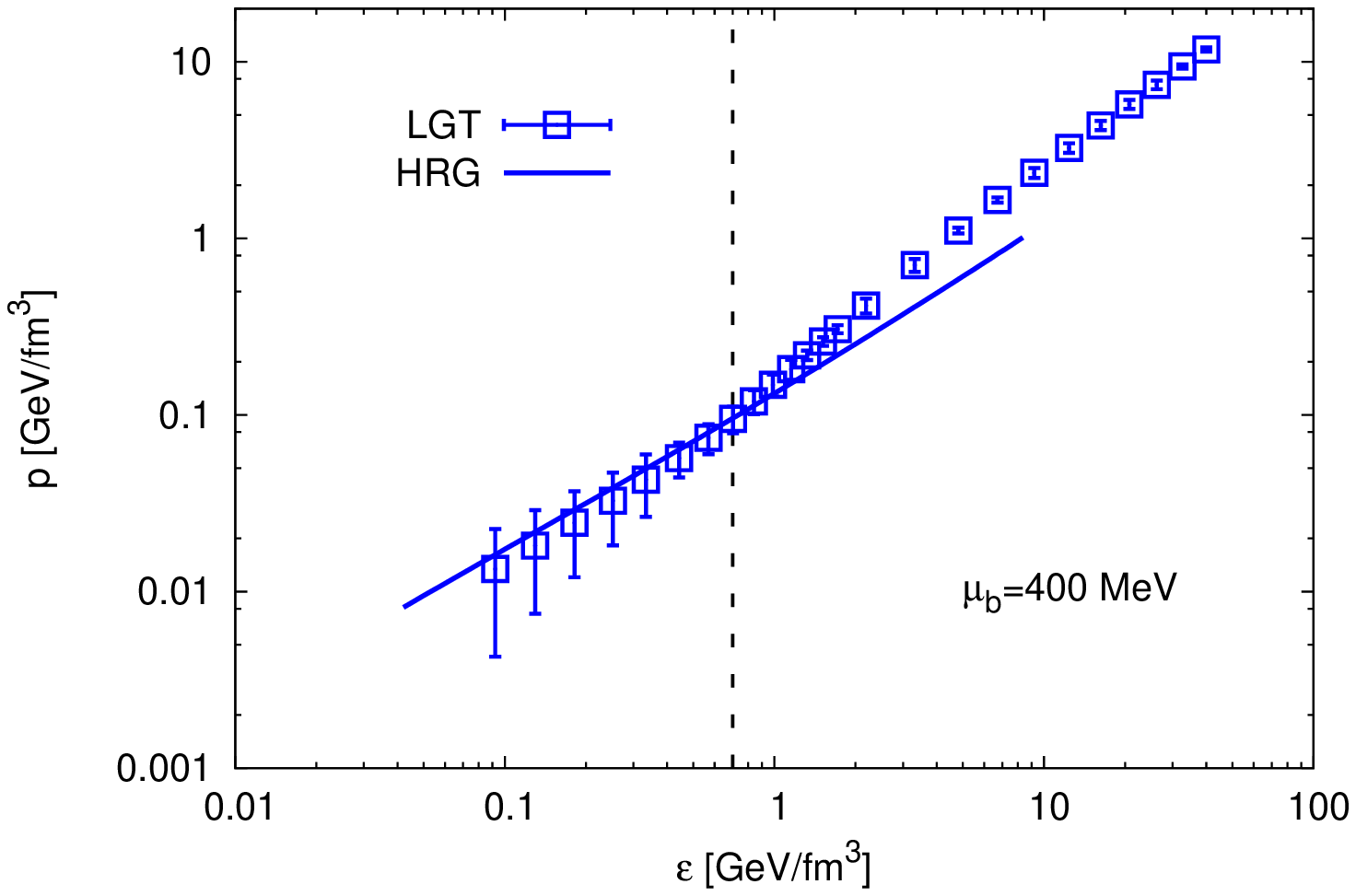}
\includegraphics[angle=-90,width=8.cm]{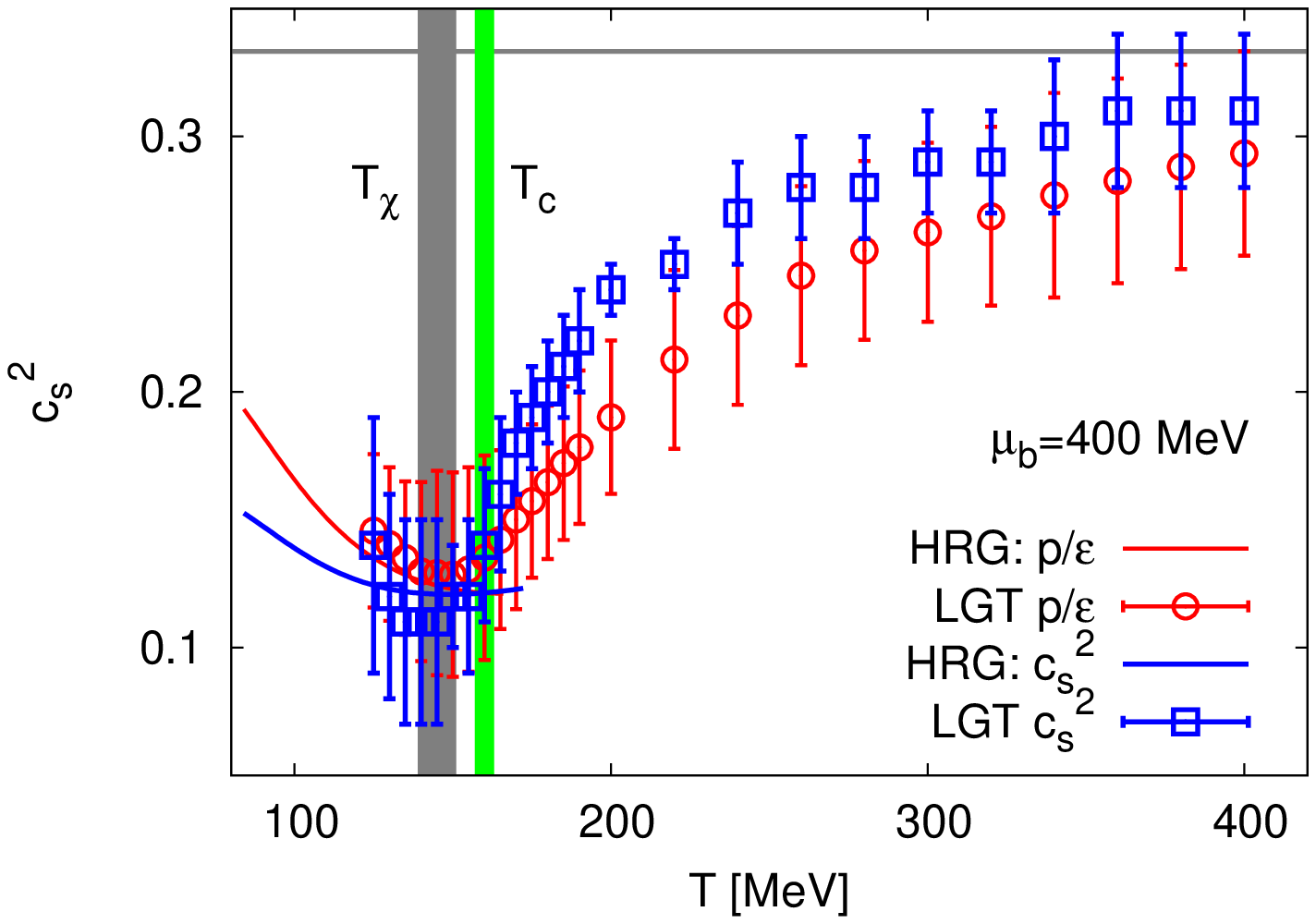}
\caption{The same as in Fig. \ref{fig:cs2lqcd1}, but at the baryon chemical, $\mu_b=400\,$MeV. The critical values are marked by the vertical strips.  Beyond these marks, HRG is not longer valid.
\label{fig:cs2lqcd2} }
\end{figure}

In Fig. \ref{fig:cs2lqcd2}, we make a comparison similar to the one given in Fig. \ref{fig:cs2lqcd1}, but at finite baryon chemical potential, $\mu_b=400\,$MeV. As discussed in section \ref{sec:mdl}, the strangeness chemical potential $\mu_S$ has to be evaluated in dependence on $\mu_b$ and $T$ in order to guarantee the conservation of the overall strange quantum numbers  in the strong interactions. It is obvious that the speed of sound squared seems to be sensitive to the dense medium. Also, it is apparent that the difference between the two expressions (\ref{eq:cs2a1}) and (\ref{eq:cs2a2}) becomes large in  dense medium. It seems that $c_s^2$ resulting from the second expression gets suppressed when increasing baryon chemical potential $\mu_b$, faster than the one from the first expression. This can be understood from the observation given in the left panel, where the barotropic relation of the pressure $p$, i.e. the equation of state, does not change  when increasing $\mu_b$ from $0$ to $400\,$MeV. %, left panel of Fig. \ref{fig:cs2lqcd1}.

It is worthwhile to notice that the HRG model seems to reproduce the lattice results on $c_s^2$ at low temperatures and finite chemical potential. Up to $T\sim 160\,$MeV, the agreement seems to be excellent. Once again, at higher temperature, a small discrepancy appears indicating that the lattice results increase, while the HRG ones slightly decrease or even remain constant. Such a discrepancy can be understood as the critical region  is apparently moved to low temperatures, $\sim 160-165\,$MeV. The QCD phase diagram can be described as follows. Increasing chemical potential leads to decreasing $T_c$ \cite{Tawfik:2004sw}. Relative to the results at vanishing $\mu_b$, the dip observed in lattice data is positioned at a lower temperature, $\sim 140-150\,$MeV, which seems to be very close to $T_{\chi}$. The critical temperature and energy density are marked by the vertical bands.

\begin{figure}[htb]
\includegraphics[angle=-90,width=10.cm]{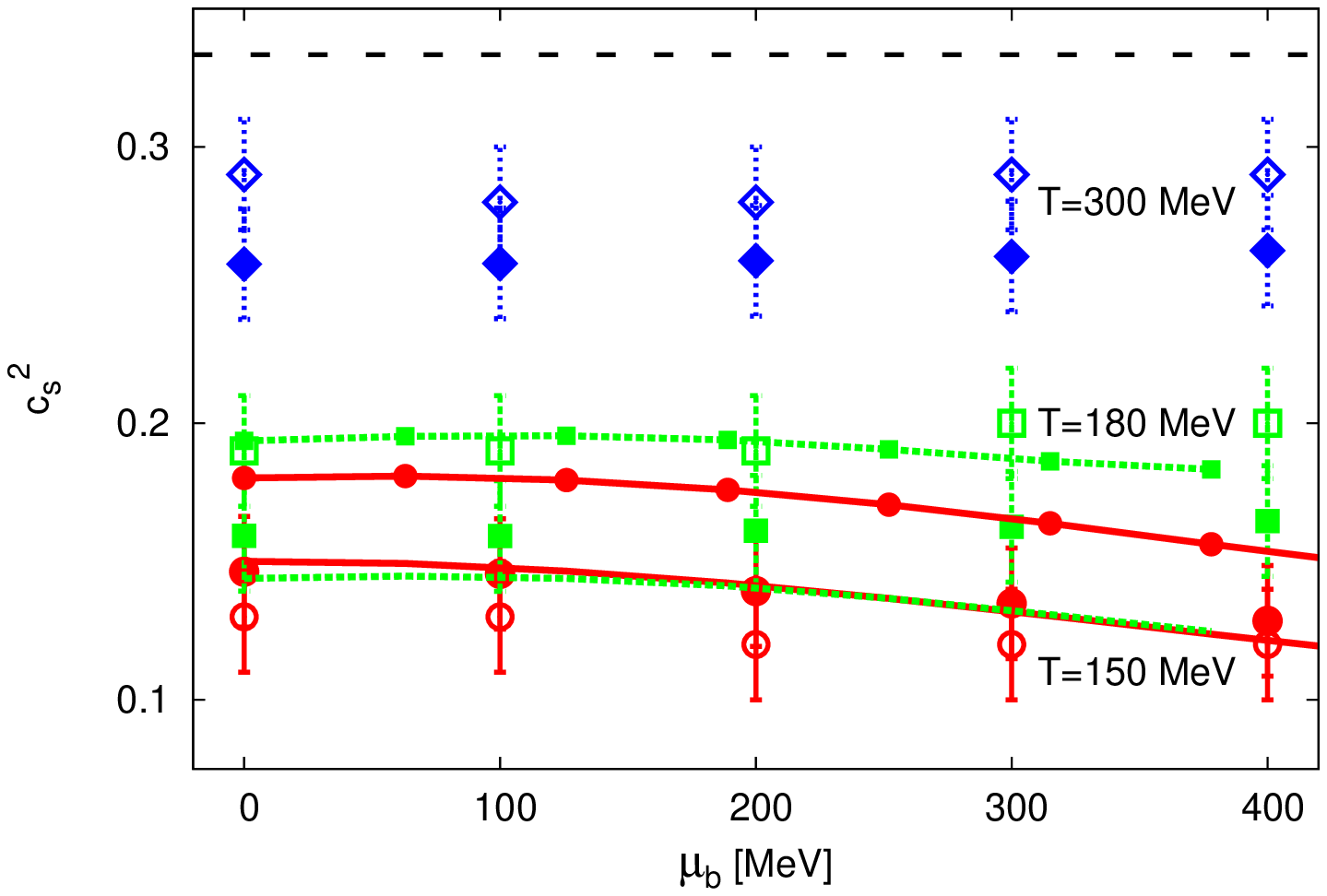}
\caption{The dependence of $c_s^2$ on the baryon chemical potential $\mu_b$ deduced in the lattice QCD simulations (symbols with error bars) \cite{lqcd4} at $T=150$, $180$ and $300~$MeV is compared with the HRG (lines and point-lines) at $T=150$ and $180$. 
\label{fig:compmub1} }
\end{figure}

At three different fixed temperatures ($150$, $180$ and $300~$MeV), the dependence of $c_s^2$ on the baryon chemical potential $\mu_b$ is presented in Fig. \ref{fig:compmub1}. The fixed temperatures are related to chiral and deconfinement phase transitions and QGP, respectively. The lattice QCD results are represented by symbols with error bars. The full (empty) symbols stand for $s/c_v$ ($\partial p/\partial \epsilon$). The HRG results, which are calculated at $T=150$ and $180$, are given by curves ($s/c_v$) and curves with points ($\partial p/\partial \epsilon$). We note that $c_s^2$ slowly decreases with increasing $\mu_b$. Furthermore, we observe that $c_s^2=s/c_v$ calculated in the HRG model at $T=180~$MeV agrees very well with the corresponding lattice QCD results. At the same temperature,  $c_s^2=\partial p/\partial \epsilon$ overestimates the lattice QCD results. The situation changes with increasing the temperature to $150~$MeV. The HRG calculations for both $c_s^2=\partial p/\partial \epsilon$ and $c_s^2=s/c_v$ agree very well with the corresponding lattice QCD simulations. The week dependence of $c_s^2$ on $\mu_b$ is observed below and above $T_c$. The only difference is that the $\partial p/\partial \epsilon<s/c_v$ below $T_c$. At $T>T_c$,  $\partial p/\partial \epsilon>s/c_v$. The HRG model confirms that $\partial p/\partial \epsilon<s/c_v$.

%%%%%%%%%%%%%%%%%%%%%%%%%%%%%%%%%%%%%%%%%%%%%%%%%%%%%%%%%%%%%%%%%%%%%%
%%%   Section IV
%%%%%%%%%%%%%%%%%%%%%%%%%%%%%%%%%%%%%%%%%%%%%%%%%%%%%%%%%%%%%%%%%%%%%%

\section{Conclusion and Outlook}
\label{sec:conl}

We conclude that the two expressions, $\partial p/\partial \epsilon$ and $s/c_v$, used in calculating the speed of sound squared are not necessarily identical. From mathematical point-of-view, they might look equivalent, as $\partial p/\partial \epsilon\equiv (\partial p/\partial T)/(\partial \epsilon/\partial T)$, which nothing but $s/c_v$. The physical nature presented in the lattice QCD simulations \cite{lqcd4} gives an estimation for such a difference. The difference between the partial and complete differentiation might partly explain this. The physical results show that these two expressions are distinguishable below and above $T_c$. Below $T_c$, $c_s^2=\partial p/\partial \epsilon$ is larger than $c_s^2=s/c_v$.  Above $T_c$, $c_s^2=s/c_v$ gets larger than $c_s^2=\partial p/\partial \epsilon$. This behavior is not affected when switching on the baryon chemical potential. At very high temperatures, both values get very close to each other and both of them approach the asymptotic value, $1/3$. In the lattice QCD simulations reported in Ref. \cite{lqcd4}, the asymptotic limit seems not be approached even at $T=400~$MeV.

When confronting the HRG results to the lattice QCD simulations,  we find that the two expressions (used to calculate $c_s^2$) are also distinguishable. Contrary to lattice QCD results, the difference calculated in the HRG model seems to get larger with increasing $\mu_b$. Furthermore, we notice that the difference decreases with increasing the temperature and almost vanishes near $T_c$. This behavior fits well with the lattice QCD results. Then, it can be concluded that the HRG results can reproduce the lattice QCD simulations, at finite $\mu_b$.

As discussed, one possible interpretation for the difference between the results of $c_s^2=\partial p/\partial \epsilon$ and $c_s^2=s/c_v$ would be related to the fluctuations associated with the specific heat. The specific heat contains several types of energy density susceptibilities, fluctuations and multiplicities \cite{cvfluct}. That both results seem to reflect such a difference might mean that these collective phenomena would be present in the HRG model. Another supportive evidence has been reported in Ref. \cite{Tawfik:2012si}.
At high $T$, formation of resonances is only achieved through strong interactions, the effective mass is assumed to approach the physical one and the strong interactions are conjectured to be taken into consideration through including heavy resonances. All these essential aspects are likely lay on collective phenomena and/or long range correlations. As the strong interaction is conjectured to be implemented, it turns to be crucial to implement other types of interactions and long range correlation functions. As a starting point, the effects of excluded volume and van der Waals repulsive interactions have to studied. In doing this, the modification from the collisionless, uncorrelated, {\it ideal} gas should be treated in framework of Uhlenbeck-Gropper approach \cite{uhlenb}. Furthermore, the different energy density fluctuations contributing to the specific heat have to be analysed, carefully. The viscous properties in Hagedorn fluid \cite{Tawfik:2010mb} would play a crucial role in bringing HRG $c_s^2$ very close to the lattice QCD one. Using different actions and lattice configurations, the dip appearing in the hadronic phase has to be calculated carefully in full lattice QCD simulations. It is obvious that the dip has its minimum value located at a range of temperatures close to $T_c$. That no dip appears in HRG would mean that the equilibrium distribution function should be a subject of modification \cite{Tawfik:2010pt,Tawfik:2010kz}. Last but not least, the higher moments have to be estimated for the lattice data, in order to understand its structure  \cite{Tawfik:2012si}.

We introduced a kind of a anatomic study for the HRG model. At low $T$, the pions (the lightest Goldstone bosons) seem to represent the main contributors to the $c_s^2$. At this scale, they are controlling the thermodynamics including the EoS, entirely. On one hand, we note that the HRG model can very well reproduce all thermodynamic quantities including the EoS of full QCD with physical quark masses \cite{Tawfik:2014eba,Karsch:2003vd,Karsch:2003zq,Redlich:2004gp,Tawfik:2004sw,Tawfik:2004vv,Tawfik:2006yq,Tawfik:2010uh,Tawfik:2010pt,Tawfik:2012zz,lqcd1,lqcd2,bazazev,lqcd3,lqcd4}. On the other hand, its results on $p$ vs. $\epsilon$ seem to partly estimate the lattice QCD results, especially at vanishing $\mu_b$. The agreement in reproducing $s/c_v$ at $\mu_b=400~$MeV is satisfactory. Performing $\chi^2$ fits would illustrate the good quality of such agreement. To give a systematic study, we divided the constituents of HRG into fermions and bosons and into non-strange and strange  resonances.
We find considerable differences between boson and fermion hadron resonances. Also there is an obvious difference between the values of $c_s^2$ calculated with and without strange hadron resonances. It is apparent that the thermal evolution of $c_s^2=\partial p/\partial \epsilon$ and $c_s^2=s/c_v$ gets  distinguishable values, especially for bosons at the intermediate temperatures. It is apparent that the fermionic  $c_s^2$ is smaller than the bosonic one. The latter decreases faster than the increase taking place in the earlier. Including strange resonances increases the fermionic $c_s^2$, on one hand. On the other hand, it decreases the bosonic $c_s^2$.

At fixed temperatures, the medium dependence of  $c_s^2(\mu_b)$ calculated in lattice QCD is compared with the HRG model. The temperatures, $150$, $180$ and $300~$MeV are related to chiral and deconfinement phase transitions and QGP, respectively. At all these temperatures, $c_s^2$ is almost insensitive to the chemical potential.

%  ============================
%              Acknowledgments
%  ============================
\section*{Acknowledgements}

The proceeding with the final phase of this work has been achieved during the visit of AT to the Institute of Nuclear Physics and the Astronomical Institute of the Uzbek Academy of Science in Tashkent. AT is very grateful to Prof. S. Ehgamberdiev, Prof. B. Ahmadov and Dr. Muslim Fazylov for their hospitality and extraordinary guest friendship. Furthermore, AT wants to acknowledge the fruitful and stimulating discussion with Dr. Ahmadjon Abdujabbarov. This research has been partly supported by the German--Egyptian Scientific Projects (GESP ID: 1378).

%%%%%%%%%%%%%%%%%%%%%%%%%%%%%%%%%%%%%%%%%%%%%%%%%%%%%%%%%%%%%%%%%%%%%%
%%%   Appendix
%%%%%%%%%%%%%%%%%%%%%%%%%%%%%%%%%%%%%%%%%%%%%%%%%%%%%%%%%%%%%%%%%%%%%%

\appendix
\section{Energy Fluctuations in Grand Canonical Ensemble}
\label{app:a}

In a grand canonical ensemble described by the partition function (\ref{eq:lnz1}), the number density 
\bea \label{eq:nA}
n(T,\mu) &=& \frac{1}{T}\, \sum_{i=1}^N\frac{g_i}{2 \pi}\, \int_0^{\infty} k^2\, dk\; \frac{e^{\frac{\mu_i-\varepsilon_i}{T}}}{1\pm e^{\frac{\mu_i-\varepsilon_i}{T}} },
\eea
and the energy density 
\bea \label{eq:eA}
\varepsilon(T,\mu) &=& \frac{1}{T}\, \sum_{i=1}^N\frac{g_i}{2 \pi}\,  \int_0^{\infty} k^2\, dk\; \varepsilon_i\; \frac{e^{\frac{\mu_i-\varepsilon_i}{T}}}{1\pm e^{\frac{\mu_i-\varepsilon_i}{T}} },
\eea
where $\varepsilon_i$ is the energy of $i$-th state. The non-normalized second moment (known as susceptibility) \cite{Tawfik:2012si} of this quantity is given by differentiating Eq. (\ref{eq:eA}) with respect to $\mu$. Then
\bea \label{eq:chiA}
\chi_{\varepsilon}(T,\mu) &=& \frac{1}{T^2}\, \sum_{i=1}^N\frac{g_i}{2 \pi}\, \int_0^{\infty} k^2\, dk\; \varepsilon_i \frac{e^{\frac{\mu_i-\varepsilon_i}{T}}}{1\pm e^{\frac{\mu_i-\varepsilon_i}{T}} }
\mp \frac{1}{T^2}\, \sum_{i=1}^N\frac{g_i}{2 \pi}\, \int_0^{\infty} k^2\, dk\; \varepsilon_i\; \frac{e^{2\frac{\mu_i-\varepsilon_i}{T}}}{\left(1\pm e^{\frac{\mu_i-\varepsilon_i}{T}}\right)^2 }, \\
&=& \langle \varepsilon \rangle\, T  \mp \frac{1}{T^2}\, \sum_{i=1}^N\frac{g_i}{2 \pi}\, \int_0^{\infty} k^2\, dk\; \varepsilon_i\; \frac{e^{2\frac{\mu_i-\varepsilon_i}{T}}}{\left(1\pm e^{\frac{\mu_i-\varepsilon_i}{T}}\right)^2 }, \label{eq:chiA2}
\eea
where the second term in Eq. (\ref{eq:chiA2}) is exactly the last line in Eq. (\ref{eq:cvserieslong}) multiplied by $T$.

Furthermore, the three integrals appearing in the second line of Eq. (\ref{eq:cvserieslong}) have a common fraction, so that this line can be approximately summarized as follows.
\bea \label{eq:threeints}
\int_0^{\infty} k^2\, dk\; {\cal O}\;  \frac{e^{\frac{\mu_i-\varepsilon_i}{T}}}{\left(1\pm e^{\frac{\mu_i-\varepsilon_i}{T}}\right)^2 },
\eea
where ${\cal O}\in[-\varepsilon_i,\varepsilon_i^2,-\varepsilon_i\, \mu_i]$. A similar integral seems to appear in the product of the susceptibility of ${\cal O}$ and the number density \cite{Tawfik:2012si}, 
\bea \chi_{{\cal O}}\, n_{{\cal O}} &=&
\sum_i^{N}
\pm 
\frac{g_i}{2 \pi} \frac{1}{T^2} \int_0^{\infty} k^2\, dk \, {\cal O}_i\, \frac{e^{(\mu_i-\varepsilon_i)/T}}{\left(1 \pm e^{(\mu_i-\varepsilon_i)/T}\right)^2} 
\left[ 
\frac{g_i}{2 \pi} \frac{1}{T} \int_0^{\infty} \frac{k^2\, dk}{1 \pm e^{(\varepsilon_i-\mu_i)/T}} 
\right].
\eea
The question arises now is: "What is the product of the susceptibility of ${\cal O}$ and the number density?" This can be found through the example
\bea
\chi\; n &=& \left[\langle N \rangle - \langle N^2 \rangle\right] \langle N \rangle
\eea
where the multiplication of two averages is obviously commutative. Generally, it results in summation of the product permutation of the two sequences times the inverse product of their length. Therefore,
\bea
\chi\; n &=& \langle N \rangle^2 - \langle N^2 \rangle \langle N \rangle.
\eea
It is apparent that each of these three terms, Eq. (\ref{eq:threeints}), gives a certain type of energy fluctuations or  multiplicities, while the fourth term given Eqs. (\ref{eq:chiA2}) and (\ref{eq:cv4thterm}) is obviously directly related to the energy susceptibility.

%%%%%%%%%%%%%%%%%%%%%%%%%%%%%%%%%%%%%%%%%%%%%%%%%%%%%%%%%%%%%%%%%%%%%%
%%%   References
%%%%%%%%%%%%%%%%%%%%%%%%%%%%%%%%%%%%%%%%%%%%%%%%%%%%%%%%%%%%%%%%%%%%%%

%-----------------------------------------------

%-----------------------------------------------

\end{document}